# Holographic formation of large area split-ring arrays for magnetic metamaterials

G. Q. Liang, W. D. Mao, H. Zou, B. C. Chen, J. F. Cao, Y. Y. Pu, X. W. Wen, and H. Z. Wang[*]

*State Key Laboratory of Optoelectronic Materials and Technologies, Zhongshan (Sun Yat-sen) University, Guangzhou 510275, People's Republic of China*

We theoretically demonstrate the formation of different kinds of two-dimensional split-ring arrays in both triangular and square lattices by one-step holographic interference. The slit width of the split-ring can be adjusted by proper polarization configurations. The dimension of the rings can be adjusted easily by using different wavelengths for interference, so the resonant frequency of the split-rings can be obtained in a wide range. Our theory is also proved in experiment. Our work would extend the application of holographic lithography to the fabrication of magnetic metamaterials.

**Key words:** interference, holographic lithography, metamaterials, split-ring resonator

---

[*]Author to whom correspondence should be addressed;

Electronic mail: stswhz@mail.sysu.edu.cn .

Left-handed metamaterial, suggested by Veselago in 1968 [1], is a kind of medium with simultaneously negative permittivity and permeability, and supporting fascinating electromagnetic properties such as the reversal of Snell's law, Doppler effect, and Cerenkov radiation. This prediction has attracted great attention since its first realization proposed by Pendry et. al. [2] and demonstrated by Smith et. al. [3], consisting of alternating layers of nonmagnetic metallic split-ring resonators (SRRs) for the negative permeability and continuous wires for the negative permittivity. After that, many interests were focused on the magnetic response of different kinds of SRRs, including single- or doule-rings, single- or multi-slits, and round or rectangular unit cells [4-9]. On the other hand, an array of pairs of short metal pillars in sub-micrometer size can contribute negative permeability at visible frequencies, which is a simplification of the resonator geometry with respect to the double-slit split-ring geometry used in microwaves [10]. In principle, the magnetic resonance arises from the inductor-capacitor circuit resonance of each unit cell, which is an analogy to a conventional LC circuit.

The prevalent procedure for fabricating such structures is as follows: firstly a template with an array of split-rings is obtained on a glass substrate by standard electron-beam lithography, then gold is evaporated onto the template by electron-beam evaporation, and after lifting off the template, gold SRRs are obtained [8]. The nanofabricated structures are with rounded off edges instead of ideal edges as in theory, however, the measured optical spectra come very close to the theoretical expectations [6, 8]. Hence, many different kinds of patterns like a coil with one or several slits can be used as templates for the SRRs, even their edges are rounded off. Such results give us an inspiration, i.e. the rapid fabrication of large area split-ring array (SRA) templates by holographic lithography (HL) for magnetic metamaterials. HL has become an efficient technique in fabricating large area templates for periodic arrays, such as for photonic crystals [11, 12] with perfect- [13-18] and quasi-periodicity [19, 20], compound microstructures [21-23], micro cavity arrays [24], and chiral microstructures [25]. Recently, templates of air rods in two-dimensional triangular lattice and one-dimensional air lines obtained by HL have been applied to fabricate large area magnetic metamaterials of the form of sandwich structure [26].

In this paper, we report the holographic formation of SRAs by one-step HL. Single-slit SRAs are formed in two-dimensional triangular and square lattices. In addition, double-slit SRAs and some kinds of compound microstructures embedded with SRAs are constructed in square lattices.

The single-slit SRAs can be looked as two-dimensional periodic structure with unit cells of special pattern like the letter 'C'. Such periodic structure can be further regarded as a compound lattice whose unit cell having four "atoms" arranged at the four corners of the letter 'C', and then properly connected to formed the required motif by appropriate polarization configuration (including the introduction of elliptical polarization), specified values of the threshold and the initial phases.

To form two-dimensional compound lattice by one-step HL, one kind of beam geometry is based on the choices of $G_{ij}$ (=$k_i$–$k_j$, where $k$s are the wave vectors) within a unit cell in the reciprocal space; another kind relies on the choices of $G_{ij}$ vectors unlimited to a unit cell [21]. The latter kind can be widely used for various complicated patterns in a compound lattice; therefore it is appropriate to the pattern construction for split-rings. For an elliptically polarized wave, let $E_{aj}$, and $\mathbf{e}_{aj}$ denote the $j$th wave's amplitude and unit polarization vector in the major axis; and let $E_{bj}$ and $\mathbf{e}_{bj}$ be those in the minor axis. The polarization degree is denoted by $A_j=|E_{bj}|/|E_{aj}|$, and $A_j=0$ for linearly polarized beams. The intensity distribution produced by the interference of multi- elliptically polarized monochromatic plane waves is given in ref. 21. For convenience, we further define the projection of $\mathbf{e}_{aj}$ onto the XOY plane making an angle of $\phi_{aj}$ with OX, and specify the initial phases and the intensities of all the beams to be the same. So the remaining control parameters for the interference patterns are the polarization direction $\phi_{aj}$ and the polarization degree $A_j$.

In order to form single-slit SRAs in a triangular lattice, five circumpolar side beams $k_j$ (j=1-5) coming from the same half space are used, making a common incident angle $\theta$=30° with respect to OZ. The beam geometry projected onto the XOY plane is shown in figure 1(a), which is based on the choices of $G_{ij}$ within a unit cell in the reciprocal space. Since $|G_{14}|=|G_{35}|=2|G_{12}|=2|G_{23}|=2|G_{34}|=2|G_{15}|$, the lattice constant of the triangular lattice formed by $G_{14}$ and $G_{35}$ is half of that formed by $G_{12}$, $G_{23}$, $G_{34}$ and $G_{15}$, so a compound triangular lattice having four sub simple lattices can be formed. According to the strategy discussed above, these four "atoms" in a unit cell should be arranged in the four corners of the letter "C" by appropriate polarization configuration $\phi_{aj}$ of the five linear polarized beams; then by introducing elliptical polarization to one of the beams, the relative sizes and positions of the four "atoms" can be adjusted [21], so finally they can be properly connected to form a split-ring at a specified threshold. The opening direction of the slit can be tuned by proper polarization configurations. The slit width can be controlled by the polarization degree $A_j$ and the exposure threshold.

Many different sets of polarization configurations can fulfill the targets, and accordingly the patterns of the split-rings may be different. Here, we give two examples, one is SRAs in the dark interference field for convenient recording in negative photoresist; and the other one is in the bright field for positive photoresist. A compound triangular lattice having four atoms in a unit cell arranged at the four corners of the letter "C" in the dark field can be formed by linear polarization configuration with $\phi_{a1}=20°$, $\phi_{a2}=170°$, $\phi_{a3}=220°$, $\phi_{a4}=300°$, $\phi_{a5}=240°$, and $A_j=0$ (j=1-5). In order to properly connect these four sets of atoms to form SRAs, elliptical polarization is introduced to only one of the five beams for simplicity. As an example, beam #3 is elliptically polarized by using $A_3=0.5$, the four sets of atoms are connected to form SRAs at a threshold 22.6% of the interference field, as shown in figure 1(b), with normalized slit width of 0.11λ (λ is the wavelength of the light source) and openings along the ΓK direction of the triangular lattice. By controlling the exposure threshold, SRAs with different slit width can be obtained. At a fixed threshold, e.g. as that in figure 1(b), the normalized slit width as a function of $A_3$ is given in figure 1(c), which provides an alternative parameter for forming rings with different slit widths. Another example for SRAs in bright interference field is also given, using polarization configuration $\phi_{a1}$=235°, $\phi_{a2}$=305°, $\phi_{a3}$=45°, $\phi_{a4}$=280°, $\phi_{a5}$=225°, and $A_2$=0.7, $A_j$=0 (j=1,3,4,5). After intensity filtering at threshold 35.8%, a split-ring array with slit width of 0.13$\lambda$ and openings along the ΓM direction of the triangular lattice are obtained, as shown in figure 1(d). These examples indicate the ability of HL in fabricating SRAs with different slit widths and opening directions.

For forming SRAs in a square lattice, six circumpolar side beams $k_i$ (i=1-6) unlimited to a unit cell of a square lattice are used, making a common incident angle $\theta$=30° with respect to OZ. The beam geometry projected onto the XOY plane is shown in figure 2(a). The traditional choice of four beams coming from the four corners of a unit cell can only provide a square compound lattice with mostly two sets of atoms, this is because there are only two sub simples lattices for superposition of the patterns [21]. So the choices of $G_{ij}$ vectors adopted here is unlimited to a unit cell. In this situation, the magnitude ratio between the largest and the smallest |$G_{ij}$| is greater than 2, so more than four atoms in a unit cell will appear. Nevertheless, by special polarization configurations and specified threshold values, only four atoms in a unit cell can be preserved.

As an example, the polarization configuration to preserved only four atoms in a unit cell which are simultaneously arranged in the four corners of the letter "C" is that: $\phi_{a1}$=40°, $\phi_{a2}$=65°, $\phi_{a3}$=115°, $\phi_{a4}$=335°, $\phi_{a5}$=160°, $\phi_{a6}$=345°, and $A_j$=0 (j=1-6). After introducing elliptical polarization to beam #4 by $A_4$=0.2, single-slit SRAs with slit width of 0.10$\lambda$ are formed at threshold of 75%, as shown in figure 2(b). At this threshold the slit width as a function of $A_4$ is shown in figure 2(c). When the threshold is increased to 81% and let $A_j=0$ (j=1-6), double-slit SRAs with slit width $0.10\lambda$ can be formed, as shown in figure 2(d). In principle, split-rings with different tropisms can be formed by varying the initial phases and polarization configuration, so the arrangement of the rings in an array can be of many different kinds.

As mention above, in square lattice, generally more than four atoms in a unit cell will appear if the beam geometry is unlimited to a unit cell in the reciprocal lattice. So this kind of beam configuration provides the possibility to combine SRAs with other interesting structures for multi functional photonic materials. Three kinds of compound microstructures embedded with SRAs are given here. The first kind is compound single-slit SRAs in a square lattice. The beam geometry projected onto the XOY plane is shown in figure 3(a). When using $\phi_{a1}=330°$, $\phi_{a2}=120°$, $\phi_{a3}=90°$, $\phi_{a4}=345°$, $\phi_{a5}=210°$, $\phi_{a6}=335°$, $A_j=0$ (j=1-6), and threshold of 62.5%, the structure are formed as shown in figure 3(b). Another array has a single-slit split-ring and a single dot in each unit cell, as shown in figure 3(d). The corresponding beam geometry is in figure 3(c), using $\phi_{a1}=335°$, $\phi_{a2}=120°$, $\phi_{a3}=95°$, $\phi_{a4}=75°$, $A_3=0.2$, $A_j=0$ (j=1, 2, 4) and threshold 18.8%. The final one is an array having two back-to-back overlapped single-slit split-rings and a long elliptical dot in each unit cell, as shown in figure 3(f). The corresponding beam geometry is in figure 3(e), using $\phi_{a1}=\phi_{a2}=350°$, $\phi_{a3}=\phi_{a4}=130°$, $\phi_{a5}=\phi_{a6}=0°$, $\phi_{a7}=\phi_{a8}=340°$, $A_j=0$ (j=1-8) and threshold 44.8%. Such compound arrays may provide not only the magnetic resonance, but also some other electromagnetic effects needed to be revealed by further researches.

The dimension of the rings can be adjusted easily by using different wavelengths for interference, and according to reference [7], the resonant frequency of the split-rings can be obtained in a wide range. If the interference wavelength is 355 nm [13], the resonant frequencies of the structures proposed above will be in the range of several tens of THz.

At last, in order to verify our theory, we show an experimental result in figure 4, corresponding to the theoretical result in figure 2 (b). It can be seen that some well identified split-ring patterns appear in the interference field, as indicated by blue circles, while in other area of the figure split-rings are deformed and some unexpected noise appears too. This is because the uniformity of our laser beams in experiment is not good enough, i.e. each point in the cross section of the laser beam dose not has the same intensity and phase, and the polarization of each laser beam in experiment is not strictly the same as those in theory. So only a small area of the interference field as the theory predicted is obtained, and the exact pattern of the rings is a little different from the theory. However, if one has high quality laser beams, it can believe that large area split-ring arrays can be fabricated by our proposed method.

Although our experiment result is not good, on the other hand, it implies that the experimental condition required by our theory has large accepted tolerance in forming the split-rings. So we conclude that our theory is correct and can be realized in practice.

In conclusion, we have demonstrate the formation of different kinds of two-dimensional SRAs by one-step holographic interference, including single-slit SRAs in both triangular and square lattices, double-slit SRAs and some kinds of compound microstructures with SRAs in a square lattice. Besides using photoresist to record the interference patterns of SRAs as templates, the SRRs may be directly obtained when the interference field is recorded by holographic plate embedded with metallic nano particles. So our theoretical results would extend the application of HL to the fabrication of magnetic metamaterials.

## Acknowledgement

This work is supported by the National Natural Science Foundation of China (10674183), National 973 of China (2004CB719804).

# References

[1] V. G. Veselago, Sov. Phys. Usp. **10** 509 (1968).

[2] J. B. Pendry, A. Holden, D. Robbins, and W. stewart, IEEE Trans. Microwave Theory Tech. **47** 2075 (1999).

[3] D. R. Smith, W. J. Padilla, D. C. Vier, S. C. Nemat-Nasser, and S. Schultz, Phys. Rev. Lett. **84** 4184 (2002).

[4] D. R. Smith, S. Schultz, P. Markos, and C. M. Souloulis, Phys. Rev. B **65** 195104 (2002).

[5] N. Katsarakis, T. Koschny, M. Kafesaki, E. N. Economou, and C. M. Soukoulis, Appl. Phys. Lett. **84** 2943 (2004).

[6] S. Linden, C. Enkrich, M. Wegener, J. Zhou, T. Koschny, and C. M. Souloulis, Science **306** 1351 (2004).

[7] J. Zhou, Th. Koschny, M. Kafesaki, E. N. Economou, J. B. Pendry, and C. M. Souloulis, Phys. Rev. Lett. **95** 223902 (2005).

[8] M. W. Klein, C. Enkrich, M. Wegener, C. M. Soukoulis, and S. Linden, Opt. Lett. **31** 1259 (2006).

[9] A. Ishikawa, T. Tanaka, and S. Kawata, Phys. Rev. Lett. **95** 237401 (2005).

[10] A. N. Grigorenko, A. K. Geim, H. F. Gleeson, Y. Zhang, A. A. Firsov, I. Y. Khrushchev, and J. Petrovic, Nature (London) **438** 335 (2005).

[11] E. Yablonovitch, Phys. Rev. Lett. **58** 2059 (1987).

[12] S. John, Phys. Rev. Lett. **58** 2486 (1987).

[13] M. Campbell, D. N. Sharp, M. T. Harrison, R. G. Denning, and A. J. Turberfield,

Nature (London) **404** 53 (2000).

[14] S. Yang, M. Megens, J. Aizenberg, P. Wiltzius, P. M. Chaikin, and W. B. Russel, Chem. Mater. **14** 2831 (2002).

[15] J. H. Moon, J. Ford, and S. Yang, Polym. Adv. Tech. **17** 83 (2006).

[16] Y. C. Zhong, S. A. Zhu, H. M. Su, H. Z. Wang, J. M. Chen, Z. H. Zeng and Y. L. Chen, Appl. Phys. Lett. **87** 061103 (2005).

[17] L. Wu, Y. C. Zhong, C. T. Chan, K. S. Wong, and G. P. Wang, Appl. Phys. Lett. **86** 241102 (2005).

[18] Y. K. Pang, J. C. Lee, C. T. Ho, and W. Y. Tam, Opt. Express **14**, 9013 (2006)

[19] X. Wang, C. Y. Ng, W. Y. Tam, C. T. Chan, and P. Sheng, Adv. Mater. (Weinheim, Ger.) **15** 1526 (2003).

[20] W. D. Mao, G. Q. Liang, H. Zou, R. Zhang, H. Z. Wang, and Z. H. Zeng, J. Opt. Soc. Am. B **23** 2046 (2006).

[21] W. D. Mao, J. W. Dong, Y. C. Zhong, G. Q. Liang, and H. Z. Wang, Opt. Express **13** 2994 (2005).

[22] L. Wu, Y. C. Zhong, K. S. Wong, G. P. Wang, and L. Yuan, Appl. Phys. Lett. **88** 091115 (2006).

[23] W. D. Mao, G. Q. Liang, H. Zou, and H. Z. Wang, Opt.Lett. **31** 1708 (2006).

[24] G. Q. Liang, W. D. Mao, Y. Y. Pu, H. Zou, H. Z. Wang, and Z. H. Zeng, Appl. Phys. Lett. **89** 041902 (2006).

[25] Y. K. Pang, J. C. W. Lee, H. F. Lee, W. Y. Tam, C. T. Chan, and P. Sheng, Opt. Express **19** 7615 (2005).

[26] N. Feth, C. Enkrich, M. Wegener, and S. Linden, Opt. Express **15** 502 (2007).

**Figure captions:**

Figure 1. Single-slit SRAs in triangular lattice formed by one-step HL. (a) Beam geometry projected onto the XOY plane. (b) SRAs formed in the dark interference field. (c) The slit width of the rings in (b) as a function of the polarization degree $A_3$ of the elliptical polarized beam #3. (d) SRAs formed in the bright interference field. Scale bars: equal to the interference wavelength $\lambda$.

Figure 2. SRAs in square lattice formed by one-step HL. (a) Beam geometry projected onto the XOY plane. (b) Single-slit SRAs formed in the bright interference field. (c) The slit width of the rings in (b) as a function of the polarization degree $A_4$ of the elliptical polarized beam #4. (d) Double-slit SRAs formed at a different threshold. Scale bars: equal to the interference wavelength $\lambda$.

Figure 3. Compound structures with SRAs in square lattice formed by one-step HL. (a), (c) and (e) are the beam geometries projected onto the XOY plane, corresponding to the arrays in each unit cell having two single-slit split-rings (b), a single-slit split-ring and a single dot (d), and two back-to-back overlapped single-slit split-rings and a long elliptical dot (f). Scale bars: equal to the interference wavelength $\lambda$.

Figure 4. Experimental interference field captured by ccd, corresponding to the theoretical result in figure 2 (b). Blue circles indicate well identified split-ring patterns.

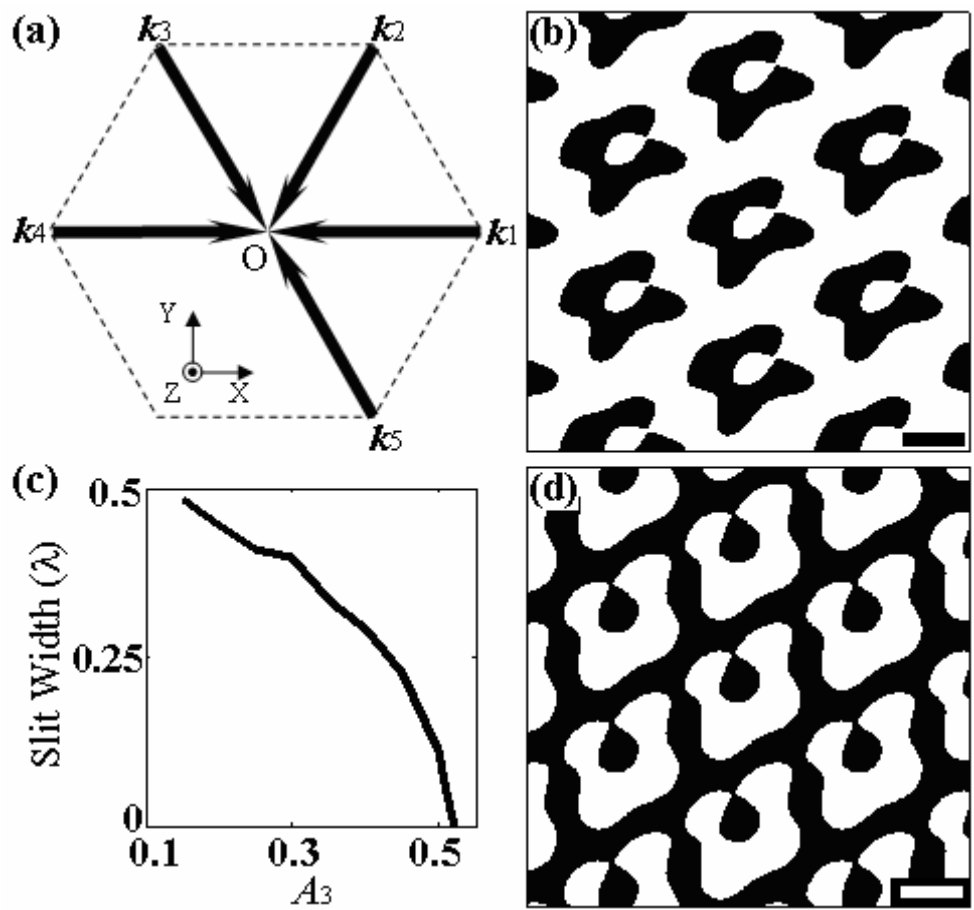

Figure 1.    G. Q. Liang and H. Z. Wang et. al.

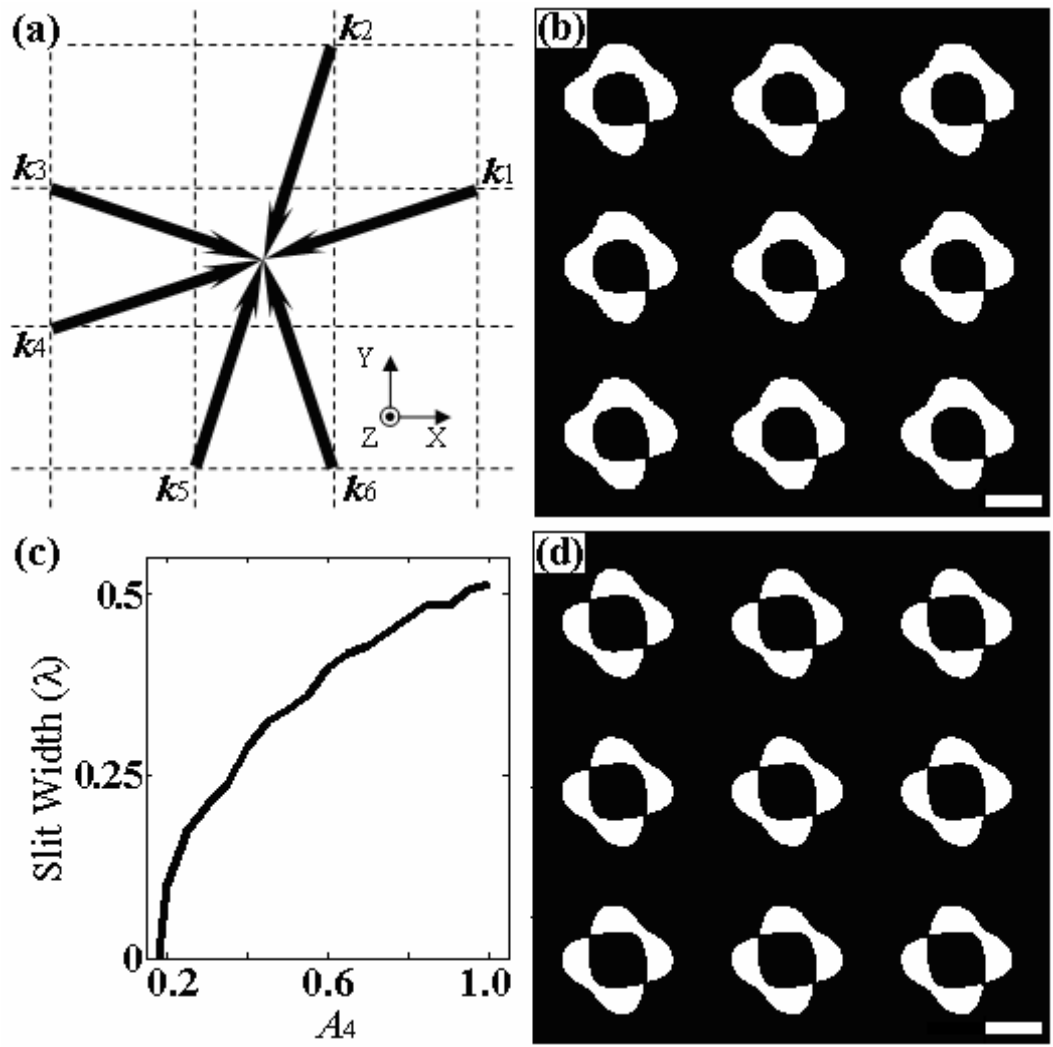

Figure 2.　G. Q. Liang and H. Z. Wang et. al.

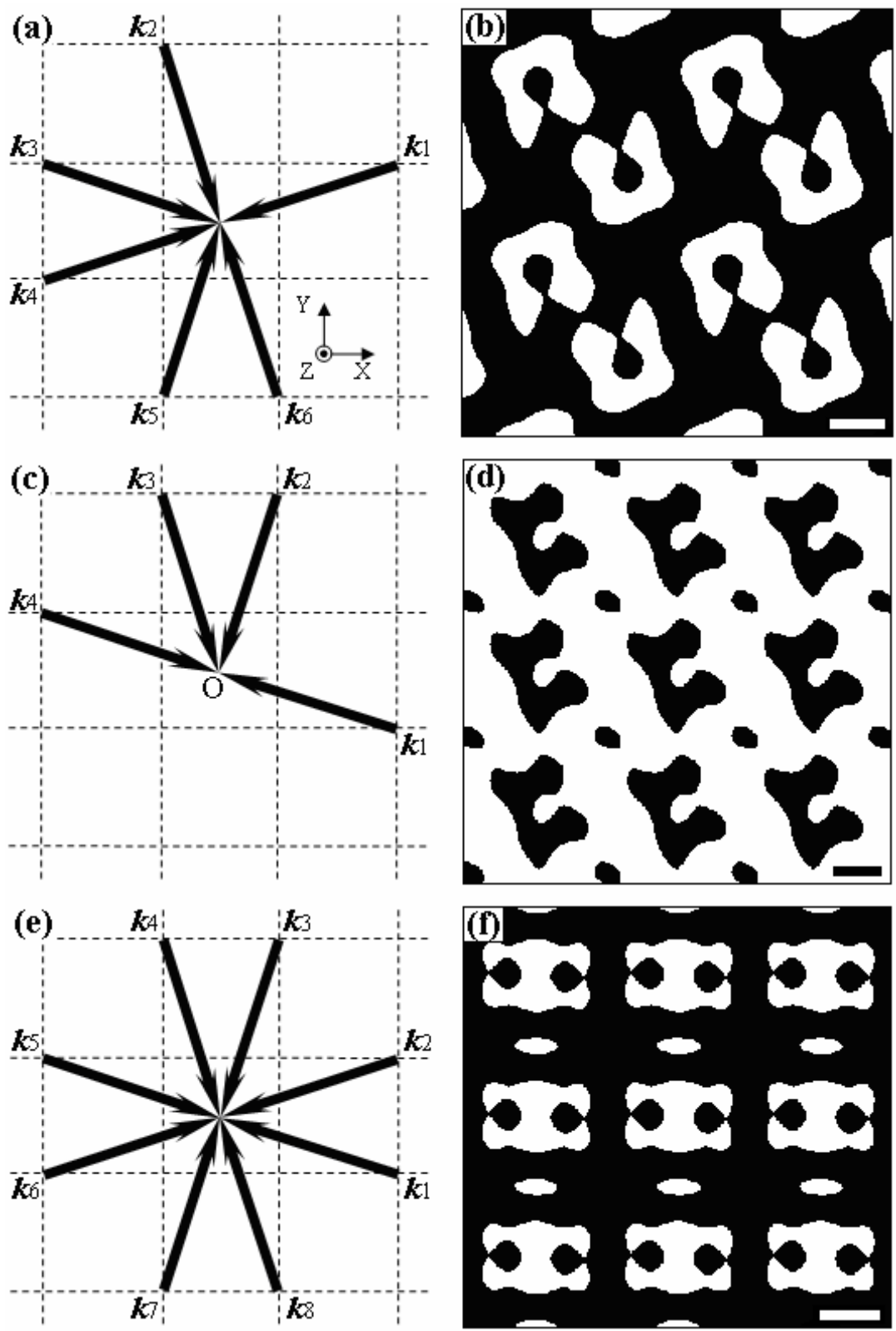

Figure 3. G. Q. Liang and H. Z. Wang et. al.

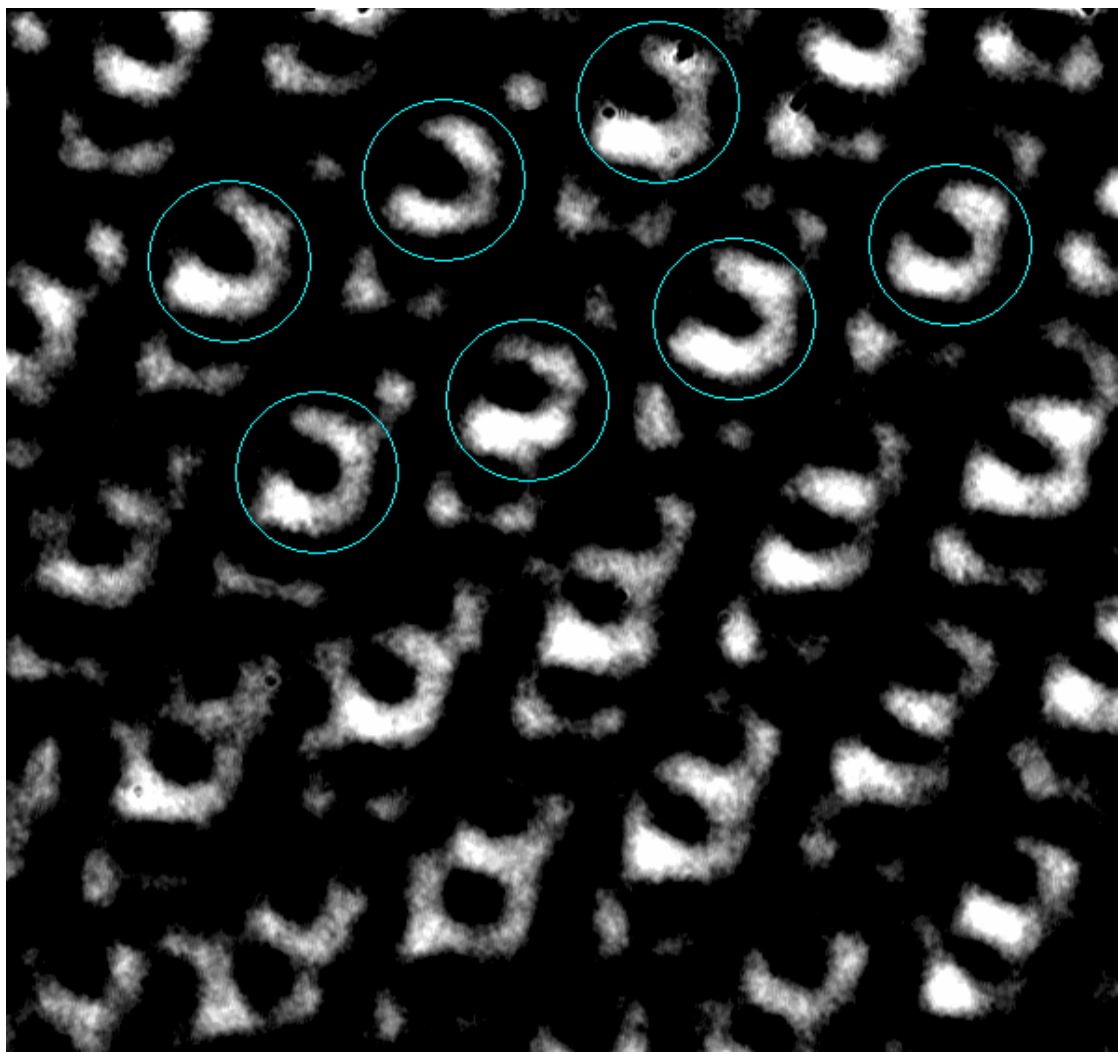

Figure 4. G. Q. Liang and H. Z. Wang et. al.